\documentclass[dvipsnames,format=sigconf,anonymous=False,review=False]{acmart}

\AtBeginDocument{%
  }

\usepackage{balance} 
\usepackage{algorithm}
\usepackage{algorithmic}
\usepackage{makecell}
\setcopyright{acmlicensed}

\begin{document}

\title{Emergence of Internal State-Modulated Swarming\\in Multi-Agent Patch Foraging System}


\author{Siddharth Chaturvedi}
\affiliation{%
 \institution{Department of Machine Learning and Neural Computing, Donders Institute for Brain, Cognition and Behaviour, Radboud University}
 \city{Nijmegen}
 \country{the Netherlands}}

\author{Ahmed El-Gazzar}
\affiliation{%
 \institution{Department of Machine Learning and Neural Computing, Donders Institute for Brain, Cognition and Behaviour, Radboud University}
 \city{Nijmegen}
 \country{the Netherlands}}

\author{Marcel van Gerven}
\affiliation{%
 \institution{Department of Machine Learning and Neural Computing, Donders Institute for Brain, Cognition and Behaviour, Radboud University}
 \city{Nijmegen}
 \country{the Netherlands}}

\renewcommand{\shortauthors}{Trovato et al.}

\begin{abstract}
Active particles sustain persistent out-of-equilibrium motion by consuming energy and can self-organize into coordinated structures such as swarms. During non-cooperative foraging under partial observability, the presence of another forager can act as a proxy signal for the nearby presence of food, promoting swarming. We validate this mechanism by simulating multiple self-propelled foragers harvesting from multiple resource patches in a continuous two-dimensional space with stochastic position updates and local passive sensing. We evolve a shared policy, implemented as a continuous time recurrent neural network that controls forager velocity, using an evolutionary strategy in which samples from the policy distribution are evaluated within the same rollout. The agents learn adaptive foraging, and when resource patches are removed, they exhibit swarming through aggregation. We further find that aggregation strength is inversely related to the amount of resource stored in a forager, consistent with risk-sensitive foraging. Analysis of the learned controller in minimal test runs reveals hidden states that are sensitive to stored resource, and clamping these states to represent lower reserves accelerates aggregation. These results demonstrate that internal state modulated swarming can emerge from local sensing alone in a multi-agent patch foraging environment.

\end{abstract}

\ccsdesc[500]{Computing methodologies~Agent / discrete models}
\ccsdesc[300]{Computing methodologies~Artificial life}
\ccsdesc[300]{Computing methodologies~Massively parallel and high-performance simulations}

\keywords{Multi-agent systems, Active particles, Patch foraging, Swarm modulation, Internal state-representation, Evolution, JAX}

\maketitle

\section{Introduction}
The emergence of coordination in the natural world, where agents compete for common resources by default, is a remarkable and complex phenomenon~\citep{CHALLET1997407}. Swarming among homogeneous groups of agents exemplifies such coordination. It is studied across various disciplines of social and natural sciences, not only for its practical applications, such as decentralized drone control~\citep{brambilla2013swarm} and crowd management~\citep{bellomo2022life, albi2019vehicular}, but also because it helps us gauge how local interaction rules and individual decision-making give rise to population-level behavior, which is reflected in the swarming pattern itself~\citep{shishkov2022social}. To understand the mechanics of such a relationship between the behaviors of individuals in a swarm and those of the emergent swarming patterns, it is useful to model them as active particle-inspired self-propelled agents.
 
Active particles refer to entities that propel themselves in their environment by consuming energy, and in doing so, they exhibit non-equilibrium dynamics~\citep{gompper20202020}. They are often used to represent flocks of birds, schools of fish, or bacterial populations in swarming simulations because they abstract away species-specific characteristics like physiology and cognition while retaining only self-propulsion, local sensing, and simple interaction rules~\citep{khadka2018active}. Since Reynolds' boids model~\citep{Reynolds87}, various models based on self-propelled agents have showcased the emergence of swarms. Traditionally, they rely on the baked-in interdependence of agents' movements with respect to each other. This is done either by directly imposing constraints on an agent's movement as a function of its neighbors~\citep{couzin2002collective, sumpter2008information, vicsek2012collective, jadbabaie2003coordination, olfati2006flocking} or by enabling communication between agents, which can be implicit (such as chemotaxis)~\citep{takata2024evolution,ramos2007computational} or explicit (such as direct messaging)~\citep{witkowski2016emergence}. Recent analytic work also derives geometric conditions under which locally interacting swarms remain cohesive without relying on communication or explicit inter-agent motion constraints~\citep{casiulis2025geometric}. Self-supervised representation learning, where agents optimize a social objective, has also been used to automatically discover swarming behavior~\citep{mattson2025discovery}. However, the existence of such minimal interdependence in a naturally competitive world is itself an assumption, and its origin remains elusive.

A logical place to probe the origins of swarming is a patch-foraging environment with active agents that only passively sense each other’s presence, without additional interaction constraints. Recent work in this direction suggests that swarming can arise because of partial observability in such an environment: a neighbor's presence may serve as an informative proxy for the presence of a hidden food source. This was realized by~\citep{loffler2023collective}, where multi-agent reinforcement-learning (MARL)-trained active colloids learned to flock while moving toward the food and to mill within the food region. Interestingly, the same policy also produced milling in the absence of a food source. In their setup, each particle had a local, occlusion-aware perception of neighbors and food within a limited field of view. A key open question is whether the presence of a neighboring agent in this context signals food availability or food depletion. Intuitively, the latter would discourage aggregation rather than promote it.

A complementary line of work under the umbrella of foraging theory predicts that decision-making is risk-sensitive and depends on an agent's internal states~\citep{mcnamara1986common}. The asset-protection principle states that well-provisioned agents are more risk-averse compared to starved agents, who are more likely to opt for risky options to prevent shortfall~\citep{moschilla2018state}. In the context of foraging agents, this can translate to resource-starved agents tolerating crowding in the form of aggregation to reduce resource variance and search time, whereas well-fed agents can repulse or remain agnostic to each other. There are also various accounts of neural mechanisms, such as evidence accumulation~\citep{hayden2011neuronal} and urgency-gating~\citep{carland2019urge} that underpin the decision-making in such foraging-related tasks. 

In this work, we aim to demonstrate the emergence of such behavior in a computational model of active particle-inspired foraging agents. We also aim to empirically analyze the learned controller behavior responsible for controlling these agents, and link it to a neural mechanism observed in natural contexts of similar settings.
In the next section, we present our model in detail, extending~\citep{loffler2023collective} by allowing agents to sense the exact proximity and resource content of entities along their line of sight while preserving occlusion. We also augment the observation of each agent with relevant internal state variables to facilitate the emergence of internal state-sensitive behavior. A single shared policy in the form of a continuous-time recurrent neural network (CTRNN)~\citep{beer1995dynamics} is trained with covariance matrix adaptation evolutionary strategy (CMA-ES)~\citep{hansen2016cma}, wherein, at each generation, we evaluate multiple policy samples concurrently in the same rollout by assigning each sample to a different forager, thereby capturing interaction effects during selection. The Results section then demonstrates adaptive foraging and internal state-dependent aggregation, where the mean spacing decreases as the amount of resource within a forager declines, aligning with the general theme of risk-sensitive foraging. We also report targeted ablations and an empirical analysis of the learned CTRNN hidden states. Crucially, in minimal two-agent test runs, when we clamp these hidden states to represent lesser stored-resource values, the onset of aggregation behavior is hastened, which is consistent with an urgency-gating-like mechanism. We conclude with limitations of our approach and future directions. The entire system is implemented in ABMax~\citep{chaturvedi2025abmax} -- a JAX-based~\citep{jax2018github} agent-based modeling framework that enables vectorized simulation and just-in-time compiled training for efficient experimentation.

\section{Model}
\begin{figure*}[h]
  \centering
  \includegraphics[width=0.85\linewidth]{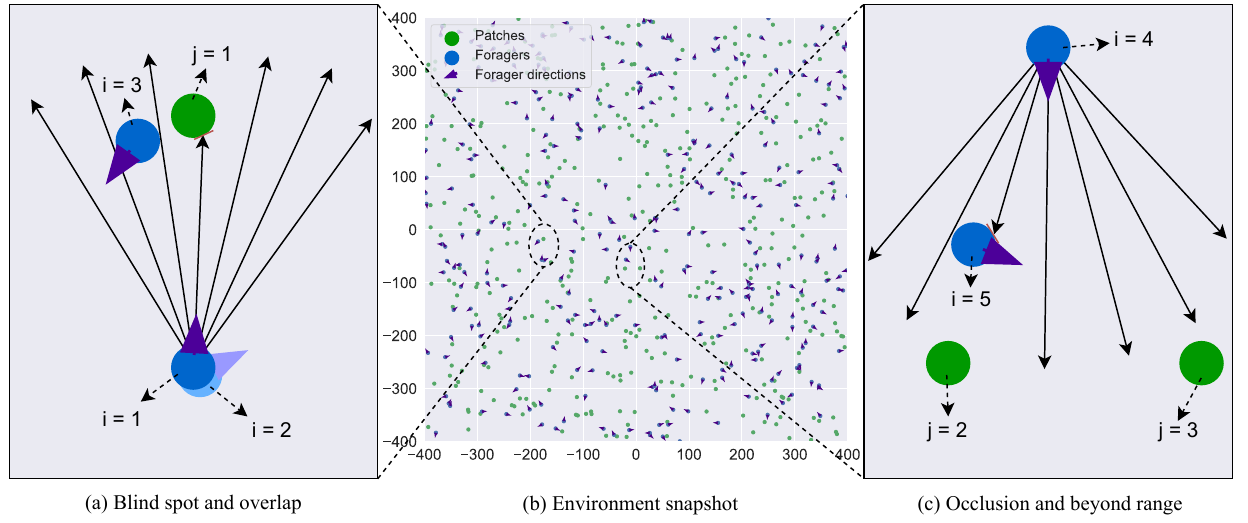}
  \caption{Environment snapshot, and local sensing limitations in the patch-foraging task.
  (b) Snapshot from one rollout with $n=300$ foragers and $m=400$ resource patches. Arrows indicate foragers' directions of gaze and movement.
  (a) Close up: forager $i=1$ emits equiangular rays; because its center lies within forager $i=2$, that forager is not registered; patch $j=1$ is detected by a ray, while forager $i=3$ falls in a blind spot between rays. 
  (c) Close up: patch $j=2$ is occluded by forager $i=5$ relative to forager $i=4$; patch $j=3$ lies beyond the ray range.}
  \Description{Three panels illustrate the environment and sensing. Panel (b) shows the full arena with many circular foragers and circular resource patches during one rollout (300 foragers, 400 patches). Panel (a) zooms into a region with three foragers and one patch. Forager i=1 sits at the center, emitting a fan of evenly spaced rays. Because its center lies inside forager i=2, that neighbor is excluded from sensing and does not register as an intersection. One ray intersects resource patch j=1, so that patch is detected. A third forager, i=3, lies between two rays and is not detected, illustrating a blind spot created by finite ray resolution. Panel (c) zooms into another region to show two more visibility limits of forager i=4: patch j=2 is directly behind forager i=5 relative to i=4 and is fully occluded; patch j=3 is farther than the maximum ray length and is therefore out of range. Together, the panels demonstrate how occlusion, blind spots, and finite sensing range shape what a forager can perceive.}
  \label{fig:environment}
\end{figure*}

The model used in this work is integrated in discrete steps where time is discretized as $t_k = k\Delta t$, $k$ being the step number and $\Delta t$ being the step size. Each run of the model is $T$ steps long. It consists of $\mathcal{A}=\{1,\dots,n\}$ foragers, each emitting $\mathcal{R} = \{1,\dots,r\}$ rays and foraging from $\mathcal{S}=\{1,\dots,m\}$ resource patches. Both foragers and resource patches are modeled as circular disks embedded in a two-dimensional Cartesian plane. The foragers can move around in the plane while the resource patches remain spatially stationary. Entities can pass through each other without a hard-body collision. A snapshot of the environment can be seen in Figure~\ref{fig:environment}(b). Model updates are organized into the following submodels. 

\subsection{Position model}
The position of the center of a forager $i\in\mathcal{A}$ at step $k\in[0,T]$ is represented in Cartesian coordinates by $\mathbf{q}_{k,i}\in \mathbb{R}^2$ and its orientation is given by $\theta_{k,i}\in(-\pi,\pi]$. For notational convenience, we drop the agent index $i$ unless necessary. Positions are constrained to a bounded space, $\|\mathbf{q}_k\|_2 \le q_{\max}$. The position states are updated as
\begin{align}\label{eq:position_update}
    \mathbf{q}_{k+1} &= \mathbf{q}_{k}+\Delta{t}\cdot\mathbf{v}_k, \qquad
    \mathbf{v}_{k+1} = s_k\!\begin{bmatrix}\cos\theta_k\\ \sin\theta_k\end{bmatrix} \\
    \theta_{k+1} &= \theta_{k} + \Delta{t}\cdot\omega_k,
     \qquad \omega_{k+1} = u_k
\end{align}
where $\mathbf{v}_k\in\mathbb{R}^2$ is its velocity along the Cartesian axes and $\omega_k\in\mathbb{R}$ is its angular velocity. The quantities $s_k\in\mathbb{R}$ and $u_k \in\mathbb{R}$ are the translational and rotational speed outputs of a forager's velocity controller, respectively, and are given by
\begin{align}\label{eq:speed_update}
    s_k &= s_{max}\cdot\sigma\left(a_{k}^{(1)}\right)\cdot\left(1 + \epsilon\cdot\xi^{(s)}_k\right) \\
    u_k &= \tanh\left(a_{k}^{(2)}\right)\cdot\left(1 + \epsilon\cdot\xi^{(u)}_k\right)
\end{align}
with $\sigma(\cdot)$ being the sigmoid function (to prevent backward movement), $\epsilon$ being a positive scalar by which Gaussian noise terms $\xi^{(s)}_k , \xi^{(u)}_k\sim \mathcal{N}(0,1)$ are scaled. We set $s_{max} = 0.5d_{\mathcal{A}}$, where $d_{\mathcal{A}}$ is the forager diameter, to scale the translational motion by body size, in line with active-particle modeling conventions~\citep{fily2012athermal}. The quantities $a_{k}^{(1)}, a_{k}^{(2)} \in \mathbb{R}$ are the outputs of a CTRNN. In this submodel, we opted for predicting actions in polar form $(s_k, u_k)$ over predicting Cartesian-velocity components, followed by recovering heading direction via $\operatorname{atan2}$ to avoid discontinuities and wrap-around near $s_k\!\approx\!0$, matching the standard differential-drive kinematics used in mobile-robot modeling~\citep{siegwart2011introduction}. 

\subsection{Resource model}
The resource patches in the simulation are modeled as stationary circular disks with constant diameter $d_{\mathcal{S}}$. At a step $k$, a patch $j\in \mathcal{S}$ offers $g_{k,j} = \gamma e_{k,j}$ amount of resource to be collected by the foragers that pass through it. Here, $e_{k,j}$ is the amount of resource in the resource patch and $\gamma$ is a positive eating rate constant. In the patch $j$ the amount of resource varies as 
\begin{equation}\label{eq:resource_lost}
    e_{k+1,j} = e_{k,j}\cdot(1+\alpha) -g_{k,j}\cdot\delta_{\mathcal{A}}(\mathbf{q}_{j},\mathbf{q}_{k,\mathcal{A}})
\end{equation}
where $\alpha$ is a positive constant representing the growth rate common to all the patches, and the overlap indicator function $\delta_{\mathcal{A}}(\cdot)$ is given as
\begin{equation}
    \delta_{\mathcal{A}}(\mathbf{q}_{j},\mathbf{q}_{k,\mathcal{A}}) =
\begin{cases}
1, & \text{if } \exists\, i \in \mathcal{A}\ \text{s.t.}\ d(\mathbf{q}_{j},\mathbf{q}_{k,i}) < \tfrac{1}{2} d_{\mathcal{S}},\\
0, & \text{otherwise.}
\end{cases}
\end{equation}
with $d(\mathbf{q}_{j},\mathbf{q}_{k,i})$ being the Euclidean distance between the position $\mathbf{q}_{j}\in\mathbb{R}^2$ of the center of the patch $j$ and the forager center $\mathbf{q}_{k,i}$ at $k$. In practice, $e_{k,j}$ is clipped between a small positive value $e_{min}$ and $e_{max}$. In the forager $i$ the amount of resource $e_{k,i}$ at $k$ evolves as
\begin{equation}\label{eq:resource_gain}
    e_{k+1,i} = e_{k,i} + \sum\limits_{j\in\mathcal{S}}\frac{g_{k,j}\cdot\delta_i(\mathbf{q}_{j},\mathbf{q}_{k,i})}{p_{k,j}} - \eta\cdot\left(\frac{|s_{k,i}|}{s_{max}} +|u_{k,i}|\right) -\mu
\end{equation}
where the second term in the equation is similar to the one used in Equation~\ref{eq:resource_lost} and represents the net amount of resource collected by the forager at $k$. Here, the indicator function $\delta_i(\mathbf{q}_{j},\mathbf{q}_{k,i})$ is 1 if the patch $j$ and forager $i$ overlap (if $ d(\mathbf{q}_{j},\mathbf{q}_{k,i}) < 0.5 d_{\mathcal{S}}$), else it is 0 and $p_{k,j}\in\{1,\dots,n\}$ represents the number of foragers present in patch $j$ at $k$. The summation is needed to account for the case in which a forager overlaps with multiple resource patches simultaneously. The third term represents the amount of resource lost by a forager as a metabolic cost, which is set proportional to the translational and rotational speeds of the forager. Here, $\eta$ is a scaling constant, and $\mu$ is a constant metabolic cost incurred by all foragers at each step. Taken together, Equation~\ref{eq:resource_lost} and~\ref{eq:resource_gain}, implement single-serve patch consumption: at step $k$ a patch $j$ loses a fixed amount of resource $g_{k,j}$, whenever at least one forager overlaps with it, while that amount is equally shared by all the $p_{k,j}$ foragers that overlap with it. This is motivated by the scramble-competition assumption prevalent in social foraging models~\citep{stephens1986foraging}.

\subsection{Controller model}
The controller for each forager $i$ is modeled using a CTRNN, which outputs its translational and rotational speeds. As input, it firstly accepts the information gathered by all the $\mathcal{R}$ equiangular rays emitted by the forager's center $\mathbf{q}_{k,i}$ of maximum length $d_{\mathcal{R}}$, spanning an angular aperture of $\pm\Delta\theta$ about its heading $\theta_{k,i}$. Each ray $\ell \in \mathcal{R}$ gathers three channels of information $\mathbf{o}_{k,\ell} = \left[o_{k,\ell}^{(1)}\,\, o_{k,\ell}^{(2)}\,\, o_{k,\ell}^{(3)}\right]^\top$ about the object $o\in\mathcal{A}\cup\mathcal{S}\setminus\mathcal{M}$ whose surface it intercepts first, where, $\mathcal{M}\subset\mathcal{A}\cup\mathcal{S}$ collects objects whose surfaces contain the ray origin itself (including the forager $i$). Thus, the ray cannot intercept their surfaces. Ray–object intersections are computed via the standard ray–circle (ray–sphere) quadratic test~\citep{ericson2004real}. The three channels of information gathered by a ray $\ell$ are given as
\begin{align}
    o_{k,\ell}^{(1)} &= \min(d(\mathbf{q}_{k,i},\mathbf{q}_{\ell,o})\,,d_\mathcal{R})\\ o^{(2)}_{k,\ell}&=\begin{cases}
        e_{k,o}&\text{if}\, o\in\mathcal{A}\land o_{k,\ell}^{(1)}<d_\mathcal{R}\\
        0&\text{otherwise}
    \end{cases}\\
    o^{(3)}_{k,\ell}&=\begin{cases}
        e_{k,o}&\text{if}\, o\in\mathcal{S}\land o_{k,\ell}^{(1)}<d_\mathcal{R}\\
        0&\text{otherwise}
    \end{cases}
\end{align}
where $d(\mathbf{q}_{k,i},\mathbf{q}_{\ell,o})$ is the Euclidean distance between $\mathbf{q}_{k,i}$ and the point of intersection $\mathbf{q}_{\ell,o}\in \mathbb{R}^2$ between $\ell$ and the surface of $o$. The amount of resource in $o$ at $k$ is given as $e_{k,o}$. Thus, in the case when no object is intersected by $\ell$, $\mathbf{o}_{k,\ell} = [d_\mathcal{R}\,0\,0]^\top$. This design choice preserves occlusion and partial-observability of the entities surrounding a forager. Figure~\ref{fig:environment}(a) depicts an instance where partial observability is realized in the model due to a gap between two rays and overlapping entities, Figure~\ref{fig:environment}(c) does the same for the cases of occlusion and limited range of the rays. 

Apart from the external observations, we also concatenate internal state of the forager $i$ given by 
\begin{equation}
    \mathbf{o}_{k,i}^{\text{istate}} = \left[\mathbf{v}_{k,i}^\top \,\,\, \omega_{k,i} \,\,\, e_{k,i} \,\,\, \Delta{e}_{k,i} \,\,\,f_{k,i}\right]^\top.
\end{equation}
Where $\mathbf{v}_{k,i}$ and $\omega_{k,i}$ are described in Equation~\ref{eq:position_update}, $e_{k,i}$ in Equation~\ref{eq:resource_gain}, $\Delta{e}_{k,i} = e_{k,i} - e_{k-1,i}$, and $f_{k,i}$ is the number of objects within which $\mathbf{q}_{k,i}$ lies (excluding forager $i$). The complete observation vector provided to the controller of the forager $i$ can be described as
\begin{equation}
    \mathbf{o}_{k,i} = [\mathbf{o}_{k,1}\,\, \mathbf{o}_{k,2}\,\,\dots\,\,\mathbf{o}_{k,r}\,\, \mathbf{o}_{k,i}^{\text{istate}}]^\top.
\end{equation}
The CTRNN controller, having hidden state size $h$, is given as
\begin{align}\label{eq:controller_update}
    &\mathbf{z}_{k+1,i} = \mathbf{z}_{k,i} + \Delta{t}\cdot\boldsymbol{\kappa}\odot( \tanh(J\cdot\mathbf{z}_{k,i} + E\cdot{\mathbf{o}_{k,i}}-\mathbf{b})-\mathbf{z}_{k,i})\\
    &\begin{bmatrix}
        a^{(1)}_{k,i} \
        a^{(2)}_{k,i}
    \end{bmatrix}^\top = D\cdot \mathbf{z}_{k,i}
\end{align}
where $\mathbf{z}_{k,i} \in \mathbb{R}^{h}$ is the vector of its hidden state, which can be considered analogous to the membrane potentials of a neuronal population~\citep{sussillo2014neural}. The inter-neuron connectivity matrix is given by $J\in\mathbb{R}^{h\times{h}}$. $E\in\mathbb{R}^{h\times{(3r+6)}}$ is the sensor-input scaling matrix and $\mathbf{b}\in\mathbb{R}^h$ is a vector of biases. Further, $\boldsymbol{\kappa} = [\tau_1^{-1}, \tau_2^{-1},\dots,\tau_h^{-1}]^\top \in\mathbb{R}^h$ is the leak-rate vector, where $\tau_c\in\mathbb{R}^+$ is the time constant of hidden state $c\in\{1,2,\dots,h\}$, operator $\odot$ represents an element-wise multiplication of two vectors. Finally, $D\in \mathbb{R}^{2\times{h}}$ is a readout matrix for the controller that maps its hidden states $\mathbf{z}_{k,i}$ to its output $a_{k,i}^{(1)}$ and $a_{k,i}^{(2)}$. The parameter set $\phi = \{J,E, \mathbf{b},\boldsymbol{\kappa},D\}$ is learned by the learning algorithm. A CTRNN model is used as the controller here, as it is a form of neural differential equation~\citep{chen2018neural} and is capable of displaying smooth dynamics that admit empirical interpretations in foraging contexts, for example as evidence-accumulation mechanisms~\citep{chaturvedi2024dynamical}.

\subsection{Foraging objective}
The aim of the foragers is to forage optimally. The optimal foraging theory~\citep{stephens1986foraging} objective calls for the maximization of the rate of resource accumulated by the forager. For a forager $i\in\mathcal{A}$, we choose controller parameters $\phi^\star_i$ to maximize terminal amount of resource gathered over a long horizon $T$:
\begin{equation}
    \phi^\star_i = \arg\max_{\phi_i}\; \mathbb{E}\left[e_{T,i}\,|\,\pi_{\phi_{i}}\right]
\end{equation}
where $e_{T,i}$ is the internal resource of forager $i$ at time $T$ when following policy $\pi_{\phi_i}$ (with parameters $\phi=\{J,E,\mathbf{b},\boldsymbol{\kappa},D\}$).

\subsection{Learning algorithm}
Learning is performed with an evolutionary strategy, specifically, the covariance matrix adaptation evolutionary strategy (CMA-ES)~\citep{hansen2016cma} is used to optimize the controller parameter set $\phi=\{J,E,\mathbf{b},\boldsymbol{\kappa}\allowbreak,D\}$. The JAX-based implementation of the algorithm from the Evosax package~\citep{lange2023evosax} is used for the purpose. We vectorize all controller parameters into a single vector $\boldsymbol{\phi}\in\mathbb{R}^{|\phi|}$ for CMA-ES. Let $\bar{\boldsymbol{\phi}}\in\mathbb{R}^{|\phi|}$ and $C_\phi\in\mathbb{R}^{|\phi|\times|\phi|}$ denote the search mean and covariance. At generation $g\in\{0,\dots,G\}$ we sample $n$ parameter vectors
\begin{equation}
    \boldsymbol{\phi}_i \sim \mathcal{N}(\bar{\boldsymbol{\phi}},\, C_\phi), \qquad i=1,\dots,n.
\end{equation}
We assign one sample per forager so that forager $i$ uses controller $\pi_{\phi_{i}}$, and evaluate all $n$ policies concurrently in the same rollout. This trick (adopted from the multi-agent water-world example in~\citep{tang2022evojax}) eliminates the need for hand-crafting a curriculum to capture the inter-forager interaction effect on the policy and speeds up the learning process by evaluating all $n$ policies concurrently under identical dynamics.
To reduce variance, we repeat this evaluation over $S$ independent seeds (distinct initial conditions for position and resource models), and define the fitness of sample $i$ as
\begin{equation}
    F_i = \mathbb{E}\left[e_{T,i}\,|\,\pi_{\phi_i}\right] =  \frac{1}{S}\sum_{s=1}^{S} e^{(s)}_{T,i}
\end{equation}
where $e^{(s)}_{T,i}$ is the terminal internal resource amount in forager $i$ at time $T$ under seed $s$. Finally, CMA-ES updates the search distribution parameters using the ranked set $\{(\boldsymbol{\phi}_i,F_i)\}_{i=1}^{n}$. The learning method is described in Algorithm~\ref{alg:cmaes}.

\begin{algorithm}[t]
\caption{CMA-ES training of CTRNN foragers (concurrent evaluation)}
\label{alg:cmaes}
\begin{small}
\begin{algorithmic}[1]
\STATE \textbf{Input:} population size $n$, generations $G$, seeds $S$, horizon $T$, initial mean $\bar{\boldsymbol{\phi}}$, covariance $C_\phi$
\FOR{$g=1$ to $G$}
  \STATE Sample $\boldsymbol{\phi}_i \sim \mathcal{N}(\bar{\boldsymbol{\phi}},\,C_\phi)$ for $i=1,\dots,n$
  \STATE Unpack each vector: $\phi_i \leftarrow \mathrm{unpack}(\boldsymbol{\phi}_i)$
  \FOR{$s=1$ to $S$}
    \STATE Initialize position and resource models with seed $s$
    \STATE Assign controllers: forager $i$ uses $\pi_{\phi_i}$
    \STATE Simulate to time $T$ using \eqref{eq:position_update}--\eqref{eq:controller_update}; record $e_{T,i}^{(s)}$
  \ENDFOR
  \STATE Compute fitness $F_i \leftarrow \frac{1}{S}\sum_{s=1}^{S} e_{T,i}^{(s)}$ for $i=1,\dots,n$
  \STATE Update CMA-ES: $(\bar{\boldsymbol{\phi}},\,C_\phi)\leftarrow \mathrm{CMAES\_Update}\,\!\big(\{(\boldsymbol{\phi}_i,F_i)\}_{i=1}^{n}\big)$
\ENDFOR
\STATE \textbf{Return} $\bar{\boldsymbol{\phi}}$
\end{algorithmic}
\end{small}
\end{algorithm}

\subsection{Simulation details}
During simulation, the entire model is stepped with an explicit forward Euler integrator of fixed step size $\Delta{t}$. The initial state values and constants used during learning are listed in Table~\ref{tab:model_training_values}. Unless otherwise stated, we use the same defaults for inference experiments. The model is implemented in the ABMax agent-based modeling framework~\citep{chaturvedi2025abmax}, built on JAX~\citep{jax2018github}, which enables batched per-agent updates and parallel evaluation across seeds, thereby reducing wall-clock training time. On an NVIDIA A100 GPU, one generation (across $S{=}8$ seeds) averaged 1.96 seconds.

\section{Experiments and results}
In this section, we present the results of the training process by commenting on the learned foraging behavior. We also conduct two inference experiments to analyze the learned swarming behavior in the absence of resource patches using the learned mean parameter vector, $\bar{\boldsymbol{\phi}}$.

\begin{figure*}[h]
  \centering
  \includegraphics[width=0.8\linewidth]{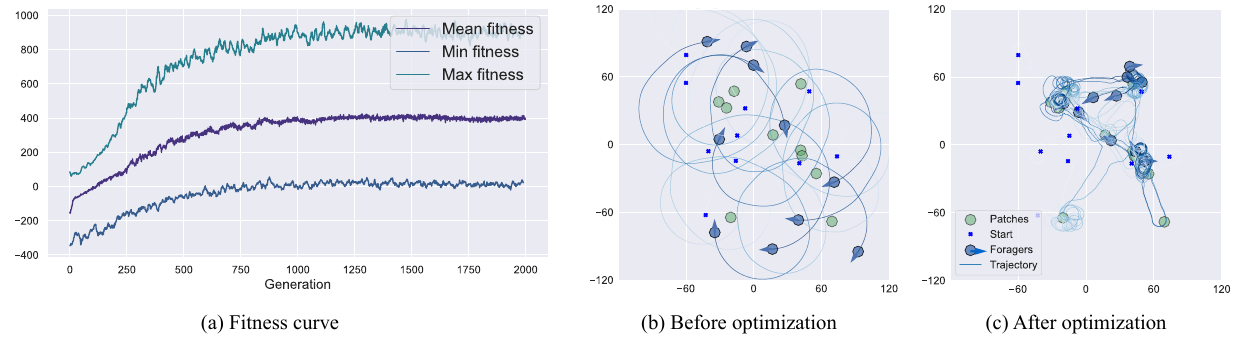}
  \caption{(a) The progression of mean, maximum, and minimum fitness across generations during evolution.
  (b) Trajectories of $n=10$ foragers in a small test environment containing $m=10$ resource patches before optimization begins simulated for $T=2{,}000$ steps. 
  (c) Trajectories of $n=10$ foragers in a small test environment containing $m=10$ resource patches with the same initial conditions as (b) after optimization, simulated for $T=2{,}000$ steps.}
  \Description{Panel (a) plots the mean, maximum, and minimum fitness against the number of generations during the optimization process. All three quantities rise monotonically and settle on a particular value, maintaining a consistent spread. Panel (b) traces out trajectories of ten foragers spawned in a small inference environment containing ten resource patches. Panel (c) traces out the trajectories for ten foragers spawned with the same initial condition as panel (b), but this time with optimized controller parameters.}
  \label{fig:fitness}
\end{figure*}

\begin{figure}[h]
  \centering
  \includegraphics[width=0.9\linewidth]{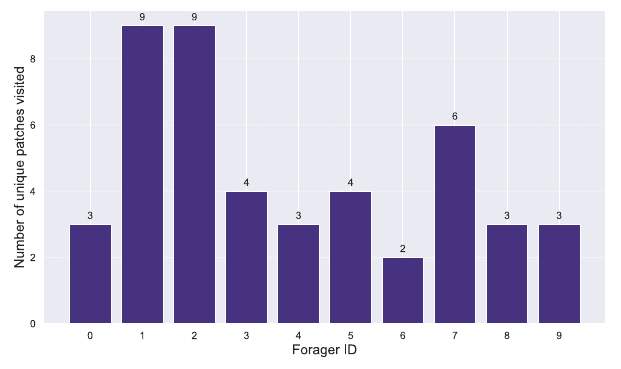}
  \caption{The number of unique patches visited by different foragers while following trajectories of Figure~\ref{fig:fitness}(c).}
  \Description{A bar graph that plots the number of unique patches visited by foragers against their ID numbers in a test run after optimization. The bars show a healthy variety, signaling that different foragers visit a different number of patches.}
  \label{fig:patch_visits}
\end{figure}

\subsection{Emergence of adaptive foraging} \label{subsec:emergence_foraging}
In Figure~\ref{fig:fitness}(a), we plot the results of the training process across generations. As can be seen, the fitness of the mean parameter vector $\bar{\boldsymbol{\phi}}$ rises and then settles to a plateau. A similar trend is visible in the values of maximum and minimum fitness scored by the samples of a CMA-ES distribution in a generation. These trends indicate a convergence of the training process. 

After training, we evaluate the learned policy in a smaller test environment with $n=10$ foragers and $m=10$ patches, spawned within a square region of length $200$. During this evaluation, all foragers share the same controller parameters set to the CMA-ES mean $\bar{\boldsymbol{\phi}}$. Figure~\ref{fig:fitness}(b) shows the forager trajectories under the untrained conditions (generation $g=0$). Here, agents move in the area, remaining largely agnostic to the resource patches and each other. Using an optimized version of the controller and the same initial conditions, yields forager trajectories (in Figure~\ref{fig:fitness}(c)) that are concentrated in and around resource patches, consistent with goal-directed exploitation.

We observe two types of foraging modes followed by the optimized policy. Firstly, a \emph{local wait-and-harvest} mode, where foragers move in a small orbit near a resource patch such that a part of the orbit lies within the patch. This effectively results in small excursions of foragers from the patch, giving the patch time to regenerate and then returning to collect the harvests. Secondly, an \emph{opportunistic traveler} mode, where foragers travel for longer distances, briefly stopping in a patch on their way to collect resources. Under this mode, foragers visit many distinct patches. These two modes are reminiscent of the dove-hawk style strategies observed in similar setups~\citep{hamon2023eco, aubert2015hunger}. Interestingly, in this model, foragers can be seen to operate in both modes, depending on their local context, such as patch occupancy, recent resource intake, or travel distance, making a clean, disjoint classification difficult. Thus, the actual foraging strategy followed by them lies in a spectrum spanned by the two modes. This spectrum is visible in Figure~\ref{fig:patch_visits}, which shows the number of unique patches visited by foragers while following the trajectories depicted in Figure~\ref{fig:fitness}(c), in the same test run. It can be observed that while some foragers visit a few patches before settling near a patch for the long term, others visit almost all the patches in the area. For a subset of time-steps in the trials, we also observed swarming-like behavior in the form of lead-follow and aggregation.

\begin{figure*}[t]
  \centering
  \includegraphics[width=0.8\linewidth]{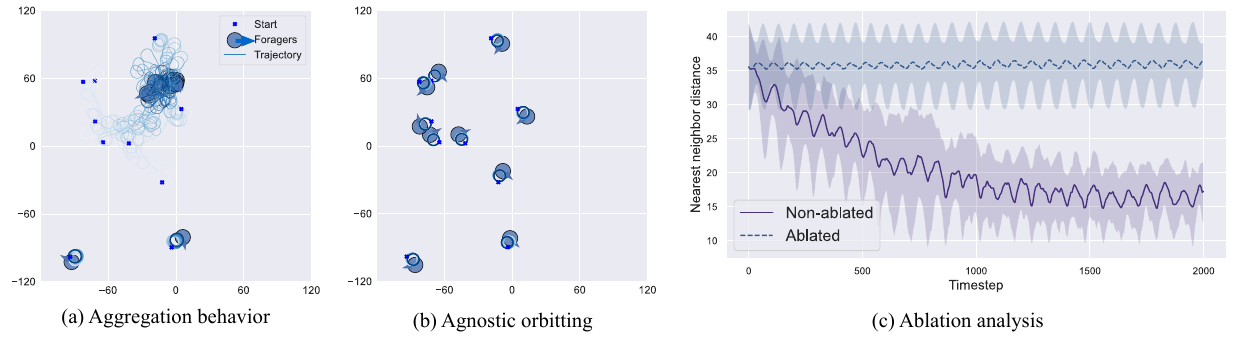}
  \caption{Swarming emerges without patches and disappears when inter-agent sensing is ablated. 
  (a) Ten foragers in a small test environment self-aggregate under the learned policy. Trajectories simulated for $T=2{,}000$ steps. 
  (b) With inter-agent sensing removed, under identical initial conditions as (a), agents remain dispersed and execute small local orbits near their spawn points.
  (c) Mean nearest neighbor distance (MNN) over time: under non-ablated (intact) conditions, it declines and stabilizes ($\approx 17$ units), whereas under ablated conditions, it stays high ($\approx 36$ units).}
  \Description{Three panels. Panel (a): overlay of 10 agent trajectories in a two-dimensional arena without resource patches. Paths curve toward a common region and densely overlap there, indicating aggregation. Panel (b): same initial setup, but the controller receives no inter-agent sensing. Trajectories form small loops around starting locations; no shared aggregation region appears. Panel (c): a line plot of mean nearest neighbor Euclidean distance versus time step, computed across the 10 agents and averaged over five random seeds. Two curves are shown: the non-ablated condition drops from about 35 units and levels near 17 units; the ablated condition remains roughly flat around 36 units. Together, the panels indicate that aggregation depends on inter-agent sensing rather than arena dynamics alone.}
  \label{fig:swarming_ablation}
\end{figure*}

\subsection{Emergence of swarming}
To probe that the observed swarming behavior while foraging was not an artifact of crowding near the resources, but emerged from the learned policy, we simulated $n=10$ foragers in a test setup (similar to Section~\ref{subsec:emergence_foraging}) without the resource patches. All foragers executed the same optimized policy $\pi_{\bar{\boldsymbol{\phi}}}$ and were initialized with internal resource $e_{0,i}=10$ for all $i\in\{1,\dots,n\}$. These test runs were simulated for five seeds representing different initial positions and orientations of the foragers. In line with the results obtained in such models~\citep{olfati2006flocking,loffler2023collective}, the foragers displayed swarming behavior in the form of aggregation (Figure~\ref{fig:swarming_ablation}(a)). To further support that swarming is rooted in sensor–motor coupling, we performed an ablation in which inter-agent sensing inputs to the controller were disabled. Specifically, for all time-steps $k$, information channels for all the rays emitted by the foragers read $[d_{\mathcal{R}} \,0\, 0]^\top$ and the entity overlap-count $f_{k,i}$ in $o^{\text{istate}}_{k,i}$ was set to $0$. In the ablated test runs, the foragers remained agnostic to each other and executed small local orbits near their spawn locations (Figure~\ref{fig:swarming_ablation}(b)). Quantitatively, Figure~\ref{fig:swarming_ablation}(c) compares the mean nearest neighbor (MNN) Euclidean distance between forager centers for the ablated and non-ablated cases, averaged over the five seeds. 

\begin{figure}[h]
  \centering
  \includegraphics[width=0.9\linewidth]{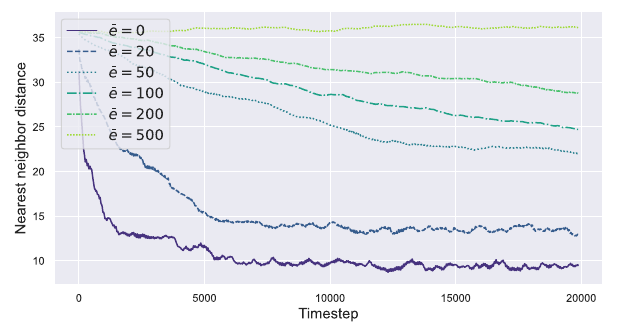}
  \caption{Progression of mean nearest neighbor (MNN) distance (calculated for $n=10$ foragers) plotted against time steps for different clamped resource levels $\bar{e}$. The plots are smoothed with a 125-step moving average filter.}
  \Description{Line chart showing the mean nearest neighbor distance (vertical axis) versus timestep (horizontal axis). Six time-series are plotted, one for each clamped forager resource level: 0, 20, 50, 100, 200, and 500. All series are averaged over five seeds and then smoothed with a 125-step moving average. The $\bar{e}=0$ curve declines the fastest and levels off at the lowest distance; as $\bar{e}$ increases, curves drop more slowly and settle at progressively higher distances. This produces a clear monotonic ordering at late times: lesser $\bar{e}$ lines lie below higher $\bar{e}$ lines, indicating stronger aggregation for lesser internal resource levels. Axis labels read “Timestep” (x) and “Nearest neighbor distance” (y). A legend labels each line by $\bar{e}$.}
  \label{fig:Mean NN dist}
\end{figure}

\begin{figure}[h]
  \centering
  \includegraphics[width=0.9\linewidth]{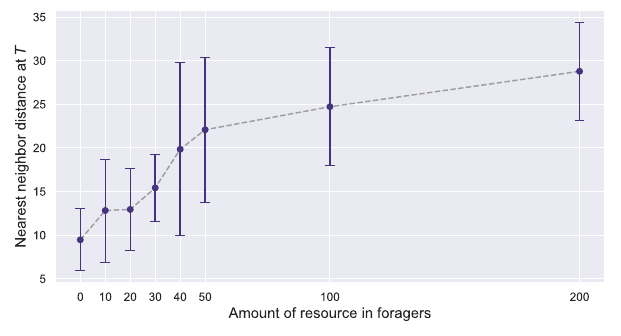}
  \caption{Final mean nearest neighbor distance versus foragers' resource level. Points show the mean across five seeds; vertical bars denote the standard deviation of $\pm1$. Higher internal resources are associated with larger inter-agent spacing (weaker aggregation). X-ticks indicate the discrete testing values, and the dashed line is for visual aid.}
  \Description{A scatter plot with error bars showing the final mean nearest neighbor distance at time $T=20{,}000$ as a function of the fixed internal resource level $\bar e$ evaluated at 0, 10, 20, 30, 40, 50, 100, and 200. The y-axis is labeled “Nearest neighbor distance at $T$.” Each marker has a vertical error bar, and markers are connected by a gray dashed line. Numeric spacing on the x-axis reflects the unequal gaps between tested resource levels. The overall trend rises: low-resource conditions correspond to smaller distances (stronger aggregation), and higher-resource conditions correspond to larger distances (weaker aggregation).}
  \label{fig:Mean NN dist vs e bar}
\end{figure}

To link swarming to internal state, we hypothesized that a forager’s internal resource value $e_{k,i}$ acts as a control parameter for aggregation. This was motivated by the risk-sensitive foraging theory arguments, wherein better-provisioned agents behave more conservatively~\citep{mcnamara1986common}. We therefore repeated the swarming test while clamping $e_{k,i}\equiv\bar{e}$ to fixed levels throughout each run, extending rollout length to $T=20{,}000$ (vs.\ $T=5{,}000$ in training). The different $\bar{e}$ levels were selected by noticing the range of $e_{k,i}$ during the training phase. The time course of MNN averaged over five seeds (again, for different initial positions and orientations) is shown in Figure~\ref{fig:Mean NN dist}, where it can be noticed that the value of MNN drops more quickly and to a smaller value for lesser $\bar{e}$. The terminal MNN at $T$ versus $\bar{e}$ is presented in Figure~\ref{fig:Mean NN dist vs e bar}. Final MNN increases with $\bar{e}$, indicating larger inter-agent spacing (lesser aggregation) at higher resource levels within the foragers.

This trend can be interpreted to be consistent with the risk-averse behavior of well-provisioned agents as proposed by the asset-protection principle~\citep{moschilla2018state}. The assumption for this link is based on the fact that, during the learning phase, the amount of resource $e_{k,i}$ in a forager $i$ at $k$ not only serves as a subset of its internal state $o^{\text{istate}}_{k,i}$, but its terminal value $e_{T,i}$ also represents the forager's fitness, which it tries to maximize. 

\begin{figure*}[h]
  \centering
  \includegraphics[width=0.8\linewidth]{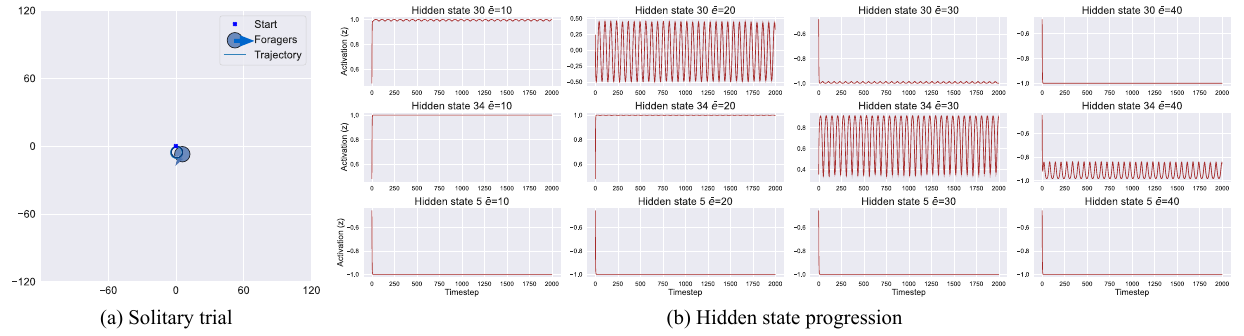}
  \caption{Single-agent CTRNN probe with clamped internal resource. (a) Example trajectory for a solitary forager with internal resource clamped at $\bar e=40$ over $2{,}000$ steps; the agent settles into a small-orbit motion primitive. \textbf{(b)} CTRNN hidden state progression (rows: state number $\in\{30,34,5\}$; columns: $\bar e\in\{10,20,30,40\}$). Hidden states $30$ and $34$ shift toward the high-saturation state as $\bar e$ decreases, whereas hidden state $5$ remains saturated near $+1$ across conditions, indicating $\bar e$-sensitive versus $\bar e$-insensitive hidden states.}
  \Description{Two panels. Panel (a): a single forager’s path in an empty arena for 2,000 steps with its internal resource fixed at 40. The path forms a tight, roughly circular loop near its spawn location, illustrating the learned small-orbit primitive in isolation. Panel (b): twelve time-series plots arranged in a 3\,$\times$\,4 grid. Rows correspond to CTRNN hidden units \#30, \#34, and \#5 (top to bottom). Columns correspond to clamped internal resource values $\bar e=10,20,30,40$ (left to right). Each trace spans time steps 0 to 2,000 on the x-axis; the y-axis is the hidden state value in $[-1,1]$. For unit \#5, all four columns show a flat line near $+1$, indicating insensitivity to $\bar e$. For units 30 and 34, the curves depend on $\bar e$: at $\bar e=40$ they remain near $-1$ or show small excursions; as $\bar e$ decreases to 30 and 20, the traces exhibit larger, earlier excursions toward higher values; at $\bar e=10$ they approach or dwell near $+1$. Overall, a lesser $\bar e$ produces earlier and larger shifts toward the high-saturation state in units 30 and 34, while unit \#5 stays saturated near $+1$ regardless of $\bar e$. All traces are means over five seeds with identical initial position and orientation.}
  \label{fig:CTRNN_free_approach}
\end{figure*}

\begin{figure*}[h]
  \centering
  \includegraphics[width=0.8\linewidth]{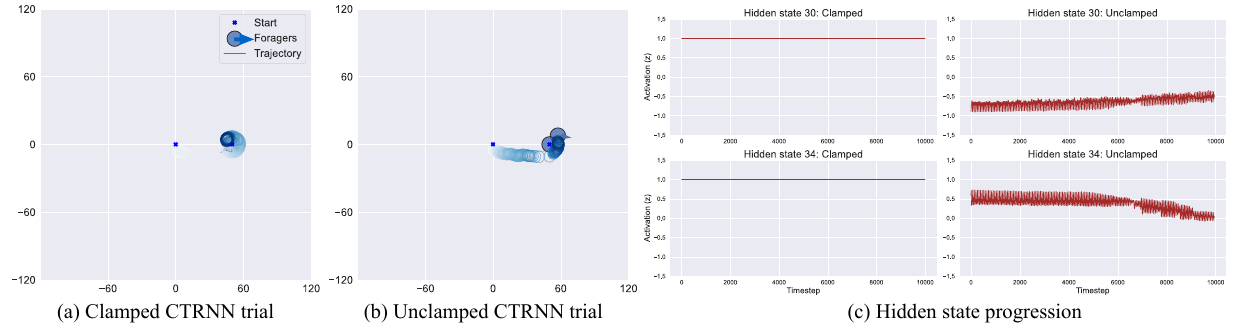}
  \caption{\textbf{(a)} Sample trajectory when the free forager’s urgency–sensitive hidden states ($30,34$) are clamped at $+1$; $\bar e_{i=1}=30$, the neighbor is fixed, $50$ units ahead with $\bar e_{i=2}=15$, and the rollout lasts $T\!=\!10{,}000$ steps. \textbf{(b)} Identical setup but with the same hidden units left free to evolve; approach occurs later. \textbf{(c)} Hidden states $30,34$ progression under the two modes. Progressions in not-clamped mode are smoothed with a $50$–step moving average for readability.}
  \Description{Three panels. Panel (a): a two–agent scene with one free forager starting at the origin and a second forager fixed $50$ units straight ahead. Internal resources are $\bar e_{i=1}=30$ (free agent) and $\bar e_{i=2}=15$ (frozen). The free agent’s CTRNN hidden units 30 and 34 are clamped at $+1$. The plotted path shows a relatively short approach followed by a tight orbit around the frozen neighbor, all within a $10{,}000$–step rollout. Panel (b): the same initial positions and parameters, except the hidden units are not clamped. The path now shows a longer initial period before approach; the final orbit forms later in time. Panel (c): a 2\,$\times$\,2 grid of line plots showing hidden–state values over time (y–axis in $[-1,1]$, x–axis from $0$ to $10{,}000$ steps). Rows correspond to units \#30 and \#34. The left column (clamped) shows flat traces near $+1$ for both units. The right column (unclamped) shows traces that begin lower and gradually increase toward $+1$, indicating a slower rise in the internal drive. All plots share identical initial pose across modes; hidden–state traces in (c) are smoothed by a $50$–step moving average.}
  \label{fig:CTRNN_clamped_not_clamped}
\end{figure*}

\begin{figure*}[h]
  \centering
  \includegraphics[width=0.8\linewidth]{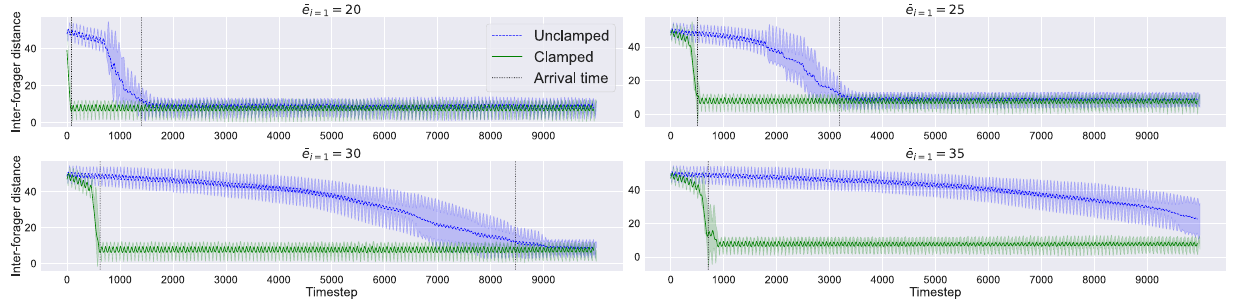}
  \caption{ Inter–forager distance versus time (mean across 5 seeds; shaded bands denote $\pm$1 s.d.) for the two-forager setup. Arrival is defined when the distance drops below the forager diameter ($10$ units). Observed arrival times (clamped vs. not-clamped) are: $86$ vs. $1411$ for $\bar{e}_{i=1}=20$, $507$ vs.\ $3192$ for $\bar{e}_{i=1}=25$, $627$ vs.\ $8476$ for $\bar{e}_{i=1}=30$, and $709$ vs.\ no arrival within $T{=}10{,}000$ for $\bar{e}_{i=1}=35$. An earlier approach under clamping across resource levels is evident.} 
  \Description{Four panels arranged in a 2-by-2 grid showing how quickly a free forager approaches a frozen neighbor under two controller conditions. Each panel plots inter–forager distance (vertical axis, units) over simulation time steps from 0 to 10,000 (horizontal axis). The two curves per panel are labeled “clamped” (the free agent’s CTRNN hidden states 30 and 34 held at +1) and not-clamped (hidden units evolve normally). Shaded regions around each curve indicate plus/minus one standard deviation across five runs with identical initial poses but different stochastic seeds. Panels correspond to different internal resource levels for the free agent: (top-left) $\bar e_{i=1}=20$, (top-right) $\bar e_{i=1}=25$, (bottom-left) $\bar e_{i=1}=30$, and (bottom-right) $\bar e_{i=1}=35$. Arrival is counted when the distance falls below 10 units and is represented using black dotted lines. Measured arrival times (clamped vs. not-clamped) are: 86 vs. 1411 in (top-left); 507 vs. 3192 in (top-right); 627 vs. 8476 in (bottom-left); and 709 vs. no arrival within the 10,000–step window in (bottom-right). The clamped condition reliably reaches the neighbor sooner at all tested resource levels.}
  \label{fig:CTRNN_dual_approach}
\end{figure*}

\subsection{CTRNN analysis}
Next, to test whether the controller learned an internal representation of a forager’s resource level, we spawned a single forager at the origin and logged the progression of its CTRNN hidden state $\mathbf{z}_{k}$ (Eq.~\ref{eq:controller_update}) while clamping its internal resource to fixed values $\bar{e}$. A sample trajectory with $\bar{e}=40$ is shown in Figure~\ref{fig:CTRNN_free_approach}(a), where, as expected, the agent executes its learned small-orbit motion primitive. We repeated this solitary trial for $\bar{e}\in\{10,20,30,40\}$ and, to account for stochasticity in the position model (Eq.~\ref{eq:position_update}), averaged the hidden states over five seeds for an identical initial pose. Figure~\ref{fig:CTRNN_free_approach}(b) plots representative CTRNN states: most resemble hidden state $5$, remaining saturated near a fixed point (approximately $\pm1$) and maintaining indifference to changes in $\bar{e}$. However, a small subset (just hidden states $30$ and $34$ for the training seed reported) exhibits monotonic shifts with $\bar{e}$. This difference in progression between the indifferent CTRNN states that occupy stable attractors and the ones that track $\bar{e}$ can be considered consistent with accounts in which a slowly varying urgency or need signal modulates the effective gain or drive of decision dynamics~\citep{carland2019urge}. Wherein, under high urgency, neuron-ensembles accumulate evidence with a higher rate reaching the decision-commitment threshold sooner in a foraging context~\citep{hayden2011neuronal,chaturvedi2024dynamical}. 

To check if the hidden states tracking $\bar{e}$ can influence the learned aggregation behavior, we augmented the test setup with another forager $i=2$, with $\bar{e}_{i=2} = 15$, fixed at an empirically selected distance of $50$ units directly along the initial heading direction of the free-to-move forager $i=1$. This setup was simulated in two modes where the two CTRNN hidden states tracking $\bar{e}$ (number $30$ and $34$) (Figure~\ref{fig:CTRNN_free_approach}(b)) for the free-to-move forager were: (i) clamped to values representing lesser $\bar{e}_{i=1}$ (specifically, $1.0$ when $\bar{e}_{i=1}=10$ (Figure~\ref{fig:CTRNN_free_approach}(b))), (ii) and free to evolve according to Equation~\ref{eq:controller_update}. The state-clamping perturbation is motivated by the causal tests of internal state variables in systems neuroscience and allows us to probe whether a need-like or urgency-like signal directly drives the approach behavior~\citep{flavell2022emergence}. The runs were repeated for different values of $\bar{e}_{i=1} \in\{20,25,30, 35\}$ and averaged over five seeds with identical pose for both the foragers to account for stochasticity in position update. As expected, the free-to-move forager approached the frozen forager at a certain time step during the test runs. 

Figures~\ref{fig:CTRNN_clamped_not_clamped}(a) and (b) show a sample trajectory in both the modes simulated for $T=10{,}000$ and $\bar{e}_{i=1} = 30$, and Figure~\ref{fig:CTRNN_clamped_not_clamped}(c) shows the progression of the CTRNN hidden states $30$ and $34$ for both these sample trajectories. Across different values of $\bar{e}_{i=1}$, the inter-forager distance (Figure~\ref{fig:CTRNN_dual_approach}) consistently converges faster in the clamped mode, compared to the non-clamped mode. Thus, the onset of the approach behavior is reliably preponed in the clamped mode where the $\bar{e}$ sensitive hidden states reflect a lesser amount of resource in the free-to-move forager. This supports a causal contribution of the learned representation of the amount of resources in a forager (exemplifying its internal state) to aggregation onset. We emphasize that the magnitude and operating range of this effect vary across training seeds and parameterizations. Mapping these dependencies and formalizing the mechanism is the subject of more rigorous analysis of the model and is deferred to future work.

\begin{table}[ht]
	\caption{Default and training parameters used for simulations}
	\label{tab:model_training_values}
    \begin{small}
	\begin{tabular}{lll}
		\text{Notation} & \text{Description} & \text{Value} \\ \midrule
		$\Delta{t}$ & Step size for simulation & $0.1$ \\
		$T$ & Length of each rollout & $5000$ \\
		$n$ & Number of foragers & $300$ \\
		$m$ & Number of resource patches & $400$ \\
		$r$ & Number of rays & $9$ \\
        $q_{max}$ & Cartesian plane limit & $10000$ \\
        $q_{0,i\in\mathcal{A}}$ & Initial forager position elements & $\sim{U(-400,400)}$ \\
        $\theta_{0,i\in\mathcal{A}}$ & Initial forager heading direction & $\sim{U(-\pi,\pi)}$ rad\\
        $\mathbf{v}_{0,i\in\mathcal{A}}$ & Initial Cartesian velocity & $[0\,0]^\top$\\
        $\omega_{0,i\in\mathcal{A}}$ & Initial angular velocity & $0.0$ \\
        $e_{0,i\in\mathcal{A}}$ & Initial resource in foragers & $\sim{U(15,30)}$\\
        $d_\mathcal{A}$ & Forager diameter & $10$ \\
        $q_{j\in\mathcal{S}}$ & Resource patch positions& $\sim{U(-400,400)}$\\
        $d_\mathcal{S}$ & Patch diameter & $10$ \\
        $e_{0,j\in\mathcal{S}}$ & Initial resource in patches & $\sim{U(0,10)}$\\
        $e_{max}$ & Maximum patch resource & $10$\\
        $e_{min}$ & Minimum patch resource & $10^{-4}$\\
        $\epsilon$ & Position noise scaling constant & $0.05$\\
        $\gamma$ & Eating rate constant & $0.3$\\
        $\alpha$ & Patch growth rate & $0.1$\\
        $\eta$ & Metabolic loss scaling constant  & $0.05$\\
        $\mu$ & Constant metabolic loss & $0.02$\\
        $\Delta{\theta}$ & Ray sensor half-span & ${\pi}/{3}$\\
        $d_\mathcal{R}$ & Max ray reach & $120$\\
        $z_{0,i\in\mathcal{A}}$ & Initial CTRNN hidden state value & $0.0$\\
        $h$ & CTRNN hidden state size & $40$\\
        - & CMA-ES distribution elite ratio & $0.3$\\
        - & CMA-ES initial step size & $0.1$\\
        $n$ & CMA-ES population size & $300$\\
        $G$ & Number of generations & $2000$\\
        $S$ & Number of seeds & $8$
        
	\end{tabular}
    \end{small}
\end{table}

\section{Discussion and conclusion}
In this work, we presented a model of many foragers acting simultaneously in a multi-patch environment under partial observability and stochastic dynamics. Given the setup, we asked two questions: (i) can coordinated behavior in the form of swarming emerge without explicit inter-agent coupling, and (ii) if so, does its strength depend on the internal state of a forager, depicted by its stored resource value. Establishing this relationship helped highlight how foragers may answer the risk-trade-off associated with swarming in such scenarios. In the spirit of active particles and ecological modeling, our aim was to minimize hand-crafted coordination rules and let complex, population-level behavior arise from local passive sensing. However, fully eliminating such assumptions is a non-trivial exercise, which is reflected in our position model as no hard-body collisions and first-order velocity dynamics over a double integrator, in our resource model as a constant growth rate and fixed positions of resource patches, and in our controller model as the range and resolution of the sensor rays and deterministic hidden state updates.

We exploited concurrent evaluation of CMA-ES policy samples in the same rollout using different foragers to capture the interaction effects during selection and accelerate training. These multi-agent interactions are key ingredients of naturalistic learning. The learned policy achieved adaptive foraging, where optimized foragers concentrated near resource patches and harvested the resources effectively. Notably, even after convergence, the spread between maximum and minimum individual fitness remained sizable, suggesting sensitivity to initial conditions, which were favorable for some configurations but not others. This hints at an ideal next step of equipping the forager controller with the ability for online adaptations~\citep{marschall2020unified}, so that they can adapt the learned policy on the fly or learn-to-learn in their local niches. 

The optimized policy expressed a combination of foraging modes spanned by a \emph{local wait-and-harvest} mode and an \emph{opportunistic traveler} mode. The foragers shifted between these modes depending on their local contexts, like patch occupancy, recent resource intake, or travel time. A definitive, disjoint distinction between the two modes was neither intended nor observed. However, making such role differentiation more explicit is a compelling avenue to explore in future work, for instance, by introducing predator-prey dynamics~\citep{yaya2023predator} within the shared controller parameter regime, or by creating different niches labeled by scarce or dense resource density in the same rollout~\citep{hamon2023eco}.

Crucially, the foragers also displayed swarming behavior in the form of aggregation in the absence of resource patches. This was validated by an ablation analysis that confirmed that the swarming was indeed due to inter-forager passive sensing rather than crowding near a food patch. Moreover, the strength of aggregation was sensitive to a forager's internal state, operationalized by the amount of resource it had. The mean nearest neighbor distance consistently increased with the amount of resource in the foragers. This modulation was in line with the risk-sensitive foraging intuitions where better provisioned foragers are more risk averse~\citep{moschilla2018state}.

\balance

Further, an empirical analysis of the CTRNN hidden states in a minimal single-agent trial revealed that a small subset of them tracked the amount of resource in the forager. Clamping them to represent depletion of resources in a two-forager setting hastened the onset of approaching behavior of the free-to-move forager towards the fixed forager. This experiment supported the existence of a causal link between the learned representation of the internal states and the swarming behavior, over correlation. Thus, the model exemplified a mechanistic bridge between the fields of neuroscience and social sciences. In the future, such empirical analysis can be complemented with more rigorous and analytical analysis represented by, for instance, a stable state analysis of the CTRNN~\citep{sussillo2013opening} or by observing the controller's trajectory in a latent space of reduced dimensions~\citep{keemink2019decoding}, to uncover the range and strength of the link.

\begin{acks}

This publication is part of the project Dutch Brain Interface Initiative (DBI$^2$) with project number 024.005.022 of the research programme Gravitation which is (partly) financed by the Dutch Research Council (NWO).
\end{acks}

\bibliographystyle{ACM-Reference-Format}
\bibliography{main}

\end{document}